# Measurements and modeling of atomic-scale sidewall roughness and losses in integrated photonic devices

*Samantha Roberts,[1,†] Xingchen Ji,[1,2,†,*] Jaime Cardenas,[3] Mateus Corato-Zanarella,[1] and Michal Lipson[1,*]*

[†]*These authors contributed equally: Samantha Roberts and Xingchen Ji*

Dr. Samantha Roberts, Dr. Xingchen Ji, Mateus Corato-Zanarella, Prof. Michal Lipson [1]
[1]Department of Electrical Engineering, Columbia University, New York, NY, USA 10027
Corresponding Author: E-mail: ml3745@columbia.edu and xingchenji@sjtu.edu.cn

Dr. Xingchen Ji[2]
[2]John Hopcroft Center for Computer Science, School of Electronic Information and Electrical Engineering, Shanghai Jiao Tong University, Shanghai, China 200240

Prof. Jaime Cardenas[3]
[3]The Institute of Optics, University of Rochester, Rochester, NY, USA 14627

**Abstract**

Atomic-level imperfections play an increasingly critical role in nanophotonic device performance. However, it remains challenging to accurately characterize the sidewall roughness with sub-nanometer resolution and directly correlate this roughness with device performance. We have developed a method that allows us to measure the sidewall roughness of waveguides made of any material (including dielectrics) using the high resolution of atomic force microscopy. We illustrate this method by measuring state-of-the-art photonic devices made of silicon nitride. We compare the roughness of devices fabricated using both DUV photo-lithography and electron-beam lithography for two different etch processes. To correlate roughness with device performance we describe what we call a new Payne-Lacey Bending model, which adds a correction factor to the widely used Payne-Lacey model so that losses in resonators and waveguides with bends can be accurately predicted given the sidewall roughness, waveguide width and bending radii. Having a better way to measure roughness and use it to predict device performance can allow researchers and engineers to optimize fabrication for state-of-the-art photonics using many materials.

**1. Introduction**

As nanophotonic devices approach losses less than 1 dB/m[1–3], atomic-level imperfections play a critical role in device performance, to a much greater degree than they do in electronics. Defects on the scale of a single atom can lead to considerable losses, even in waveguides with larger, micron-scale cross-sections. Roughness of waveguide sidewalls limits the optical performance and is typically the largest source of loss. Characterizing the magnitude of this roughness is crucial to understanding how to minimize these imperfections during fabrication. However, sidewall roughness is usually difficult to measure, especially on the sub-nanometer scale required for the state-of-the-art devices at present. It is important not only to be able to correlate the roughness to the losses of straight waveguide devices, but also to predict the losses in any photonic circuit, including those with rings and bends. Having both a direct method to measure sidewall roughness with sub-nm resolution and a model to convert this roughness to device performance permits development and optimization of fabrication processes to achieve low-loss state-of-the-art photonic devices with various materials.

Scanning Electron Microscopy (SEM) is a common method to acquire 2-dimensional (2D) images of a waveguide to infer waveguide sidewall roughness[4]. However, the resolution limit of SEM is on the order of nanometers to 10's of nanometers depending on the conductive properties of the samples[5]. Therefore, the sub-nanometer precision required for measuring roughness in state-of-the-art photonic waveguides is beyond the capability of this technique. With dielectric samples, the charging effect caused by the lack of surface conductivity and the imaging artifacts caused by focusing and stigmation offsets aggravate this resolution limitation, which make the measurement of sidewall roughness measurements susceptible to error even with sophisticated image analysis methods. Confocal microscopy imaging has also been used to image and extrapolate sidewall roughness[6], but this optical technique is also hindered by resolution limits which prevent it from measuring with sub-nm accuracy.

Atomic Force Microscopy (AFM) is a direct surface measurement technique that has atomic scale sensitivity with a resolution on the order of 0.1 nm routinely achieved, and is not limited by the material properties of the samples[7]. However, measuring the sidewalls of waveguides is not possible with standard AFM as the sidewall surface is oriented vertically with respect to the AFM tip, and thus is unable to be accessed conventionally. At present, there are specialized AFMs designed with a rotated scanning axis[8,9], which can implement customized AFM tips to gain access to the sidewalls, but these are not standard in most laboratories. One technique to circumvent this problem is to remove the waveguide from the surface by wet chemical etching and lay it on its side so the sidewall is facing the AFM tip. This has been demonstrated for silicon waveguides[10]. However, this process relies on the material of the waveguide and the substrate, and the wet chemical etching can change the characteristics of waveguide sidewalls, which results in an incorrect roughness or correlation length. Another technique has been demonstrated by carefully cleaving the waveguide, but it requires careful preparation of the samples, challenging edge-on AFM imaging of small waveguide pieces, and is heavily dependent on the substrate[11].

We have developed a method that allows us to directly measure the sidewalls of waveguides made of any material (including dielectrics) using the high resolution of AFM. We illustrate this method by measuring state-of-the-art photonic devices made of silicon nitride ($Si_3N_4$). We chose silicon nitride as our example, because it is a superb photonic platform which combines the beneficial properties of a wide transparency range, a high refractive index, and compatibility in large-scale semiconductor manufacturing[12–15]. Due to the recent progress in realizing ultra-low losses in $Si_3N_4$[16–19], it has been used in numerous fields including battery-operated integrated frequency comb generators[20], communications[21], optical gyroscopes[22,23], spectroscopy[24,25], biological imaging[26,27] and photonic quantum circuits[28,29].

## 2. Sample Fabrication Method

We fabricate the waveguides from a 730 nm silicon nitride ($Si_3N_4$) film deposited by Low Pressure Chemical Vapor Deposition (LPCVD) at 800°C onto double-sided-polished silicon wafers with 3 μm of grown thermal oxide. Lithography is used to define the waveguide device pattern, and then plasma etching is used to transfer the pattern into the film, both steps that introduce sidewall roughness. It is important to elucidate the roughness caused by each of these steps, so as to select the optimal lithography and etching processes to minimize the waveguide sidewall roughness.

In this study, we directly compare sidewall roughness of waveguides fabricated using three different combinations of lithography and etching processes:
- Process 1: DUV photo-lithography with a standard $CHF_3/O_2$ etching recipe
- Process 2: DUV photo-lithography with a low-polymer $CHF_3/O_2/N_2$ etching recipe
- Process 3: Electron-beam lithography with a low-polymer $CHF_3/O_2/N_2$ etching recipe

We chose and compared two commonly used patterning methods, namely DUV photo-lithography and electron-beam (e-beam) lithography. Our DUV photo-lithography mask is written on a Heidelberg DWL2000 Mask Writer using a wavelength of 405 nm and developed in a wet chrome etch using a Hamatech developer tool. The mask pattern is projected onto the $Si_3N_4$ wafer using an ASML DUV stepper with 248 nm light, (ASML PAS 5500/300C DUV) where it undergoes a 4X reduction using the optics of the stepper. The waveguide pattern is transferred to the wafer using UVN30 0.5 DUV photoresist (Rohm and Haas Electronics Materials). Meanwhile, electron-beam lithography is done with a JEOL 9500 100 kV using a beam current of 2 nA to direct-write the waveguide features into a MaN-2403 (micro resist technology Gmbh) negative resist.

The patterns defined by the above methods are transferred from the resist to the waveguide nitride layer via two different etch processes. We compare a standard nitride $CHF_3/O_2$ etch recipe, which is known to leave behind a fluoride polymer residue on the sidewall with an optimized (low-polymer) etch recipe ($CHF_3/O_2/N_2$). This low-polymer etch recipe uses an increased oxygen concentration to strip the polymer from the waveguide during the etch process. An oxide mask is used to ensure successful pattern transfer and nitrogen is added to the plasma to improve the etch selectivity between the oxide and the nitride[30,31].

Once the waveguides are patterned and etched, we facilitate sidewall measurement by removing the surrounding silicon substrate to create a tall and thin silicon pillar, while keeping the waveguide fully protected. This is easily toppled onto its side and makes the waveguide sidewall readily accessible by a standard AFM. **Figure 1(a-e)** illustrates our approach of preparing the sample for measurement. The waveguide is first protected by a thick photoresist (SPR 7.0 by Rohm and Haas Electronics Materials) that covers the waveguide by 1.5 µm on all sides, protecting the sidewalls of waveguide from all subsequent processing (Figure 1(b)). After etching through the buried oxide with a $CHF_3/O_2$ reactive-ion etching, we use a Bosch deep silicon etch to remove 100 μm of the silicon substrate surrounding the waveguide. This leaves behind a 100 μm tall and 5 µm wide silicon pillar with the waveguide sitting on top (Figure 1(c)). The remaining resist is stripped by oxygen plasma treatment without changing the sidewall characteristics [32] (Figure 1(d)). The Bosch etch introduces a slight (~0.5 degree) slant to the pillar sidewall, making the base of the pillar narrower than the top. We push the silicon pillar with the waveguide onto its side using a tungsten microelectrode probe manipulated with a micro positioner. When the silicon pillar is on its side, the sidewall of the waveguide is facing upwards, and the sample can be easily accessed by a standard AFM from the top (Figure 1(e)).

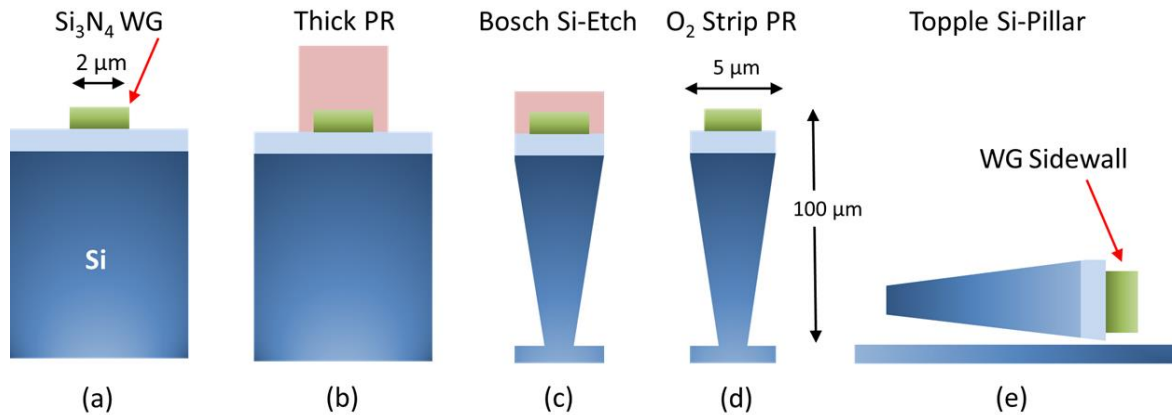

**Figure 1.** Schematic shows the process of fabricating a waveguide pillar (not to scale). (a) Linear silicon nitride ($Si_3N_4$) waveguide segments are etched on a silicon wafer with buried oxide (BOX) cladding. (b) Thick protective photoresist covers all sides of the waveguide to completely protect the sidewalls from subsequent processing. (c) The BOX and the silicon surrounding the waveguide are etched away so there is a freestanding Si pillar approximately 100 µm tall and 5 µm wide. (d) The photoresist is removed in an oxygen plasma treatment to expose the waveguide sidewalls. (e) The Si pillar/waveguide is then toppled onto its side using a tungsten microelectrode probe manipulated with a micro positioner. Once on its side the sidewall can be imaged using standard AFM techniques.

## 3. Sidewall Roughness Analysis and Results

Our sample fabrication method allows us to fully protect the waveguide during all fabrication steps, and can be used to prepare waveguides of any material to be directly measured by AFM, as long as the waveguide is on a substrate that allows a high selectivity deep etch. The measurements of the roughness and the correlation length give a description of both the magnitude of the roughness and the periodicity of its occurrence. These measurements allow us to further predict the performance of the device.

Measurements of waveguide sidewall roughness were made on a Bruker Dimension 3100 Veeco AFM configured in tapping mode. We used commercially available Nanotech ATEC-NC-20 AFM tips which feature a tetrahedral tip that protrudes from the very end of the cantilever as illustrated in **Figure 2(a)**. This geometry allows us to land the tip directly onto the sidewall, and furthermore gives unimpeded access to the waveguide sidewall. Once the silicon pillars were lying on their sides, the waveguide was located using the AFM optical microscope and was oriented so that it was perpendicular to the protruding AFM tip as seen in **Figure 2(b)**.

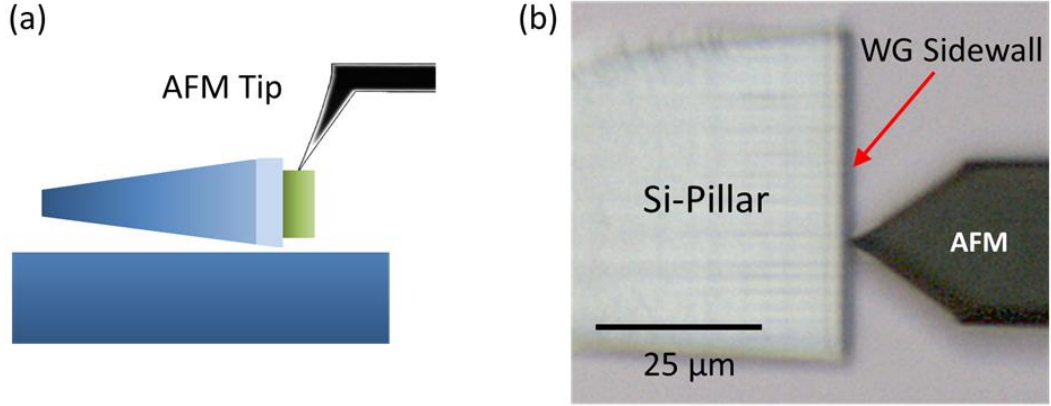

**Figure 2.** AFM imaging of waveguide sidewalls. (a) Side view schematic of an ATEC-NC-20 AFM tip (black) imaging the sidewall of a waveguide (green - not to scale). (b) Top-down optical image (60X) taken with the AFM camera of the AFM cantilever scanning the waveguide sidewall surface. The side of the Bosch etched silicon pillar is the left (white) part of this image.

We acquired AFM height images 2 µm long and 500 nm high by scanning along the propagation axis of the waveguide. A height measurement was recorded every 1-2 nm with a scan speed of 1 µm/second. Height, phase, and amplitude error data was collected for each scan. NanoScope Analysis software (Bruker v1.50) was used to visualize and analyze the scanned images and to extract the area sidewall roughness and correlation length. We began with a second order plane fit to remove the curvature effect associated with the mechanical functioning of the AFM. The height data was exported as an ASCII matrix, and read into a MATLAB script for further 1D analysis. For each row of the data, line cuts were taken along the horizontal X-axis (along the length of the waveguide) (**Figure 3(a-b)**) and the root-mean-square (RMS) roughness was calculated to find the anisotropic roughness due to the patterning and lithography, The same process was done along the vertical Y-axis to extract the isotropic roughness due to the etch process (not shown). In **Figure 3(c)**, we see the cross section of three line cuts along the horizontal X-axis for the 2 µm scan.

While the RMS value gives a measure of the vertical roughness of the sample, the correlation length, ($L_c$), gives the measure of the lateral distribution of the features. Physically, the correlation length is the horizontal distance one must move along the waveguide to observe self-similar features. To find this we compute the 1D autocorrelation function (ACF) using **Equation 1**, a plot of which can be seen in **Figure 3(d)**. This is the correlation of the surface ($h_x$) with itself in both the positive and negative directions. The peak on this graph gives the square of the root mean square (RMS) roughness ($\sigma$) and the correlation length is the distance it takes for the autocorrelation function to fall to 1/e of its peak value.

$$R_x(\tau) = \langle h_x h_{x+\tau} \rangle = \sigma^2 \, exp\left(-\frac{|\tau_x|}{L_c}\right) \quad (1)$$

In the frequency domain, the PSD is computed by taking the Fourier transform of the autocorrelation function (**Equation 2**) where $k$ denotes the spatial frequency along the x-axis, and the inverse corner frequency is the correlation length. The PSD for the three line cuts is plotted in **Figure 3(e)**.

$$G_x(k) = FT[R_x(\tau)] = \frac{2\sigma^2 L_c}{1 + (L_c k)^2} \quad (2)$$

We use the PSD to validate that the scan length and resolution are sufficient to contain the corner frequency, and compute the actual correlation length using the decay of the autocorrelation function for each line cut in the data matrix.

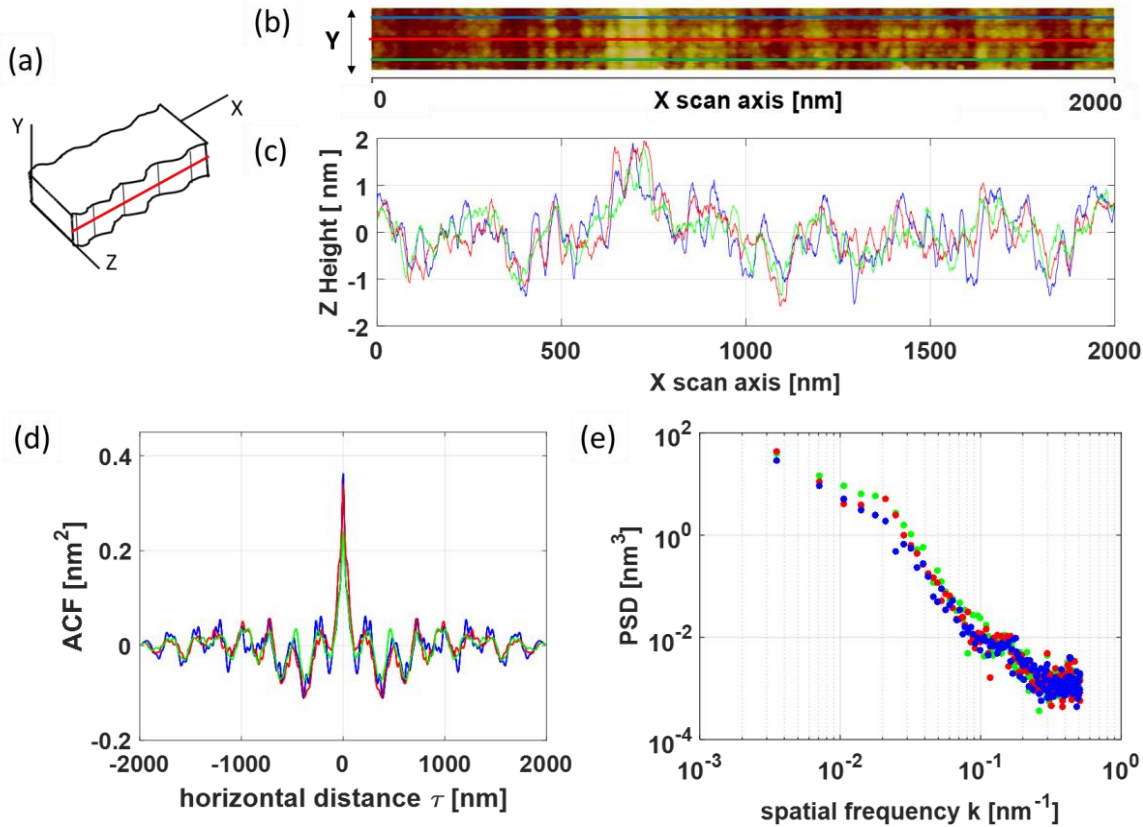

**Figure 3.** Analysis of an AFM scan of Process 3 - Electron-beam lithography with low-polymer etch. (a) A schematic denoting a waveguide and its x, y and z axes, with a red line on the sidewall showing where the scans and line cuts are taken. (b) A 2D AFM height plot showing the waveguide sidewall and the position of three horizontal line-cuts through the data. (c) The line-cuts are shown of height as a function of scan length which shows the roughness of the waveguide sidewall given by the heights (z). The blue trace is at the top of the waveguide, the red is in the middle, and the green is at the bottom of the waveguide. (d) Peak of the autocorrelation function where the maximal point gives the value of the RMS roughness squared. (e) 1D power spectral density function showing the frequency content of the scan, where the corner frequency indicates the inverse correlation length ($L_c$) and can be seen at $k \approx 2.13 \times 10^{-2}$ (nm$^{-1}$).

We used AFM to directly measure and compare samples fabricated with the three different processes described above. The results of these measurements are shown in **Table 1**. We see changing from the standard etch to the low-polymer etch reduces the RMS roughness (along the propagation axis) for DUV photolithography from 2.83 nm to 1.23 nm, and that waveguides patterned with electron-beam lithography further reduce the roughness of the low-polymer etch to 0.53 nm. For the parameters we explored, a combination of e-beam lithography with the low-polymer etch produced the smoothest sidewalls.

**Table 1.** Summary of side wall roughness and correlation length results as measured directly with AFM.

|  | Lithography | Etch Recipe | σ (nm) | σ[min-max] (nm) | Lc (nm) | Lc[min-max] (nm) |
|---|---|---|---|---|---|---|
| Process 1 | DUV-Photo | $CHF_3/O_2$ | 2.83 | 1.77 – 3.66 | 96.0 | 56.6 – 170.1 |
| Process 2 | DUV-Photo | $CHF_3/O_2/N_2$ | 1.23 | 1.07 – 1.48 | 145.6 | 61.5 – 227.5 |
| Process 3 | Electron-Beam | $CHF_3/O_2/N_2$ | 0.53 | 0.47 – 0.75 | 47.4 | 25.9 – 66.9 |

We show the three-dimensional (3D) AFM images of the three processes we measured in **Figure 4**. Each figure has the same color scale and is sized to best visualize how each process affects the magnitude of the sidewall roughness.

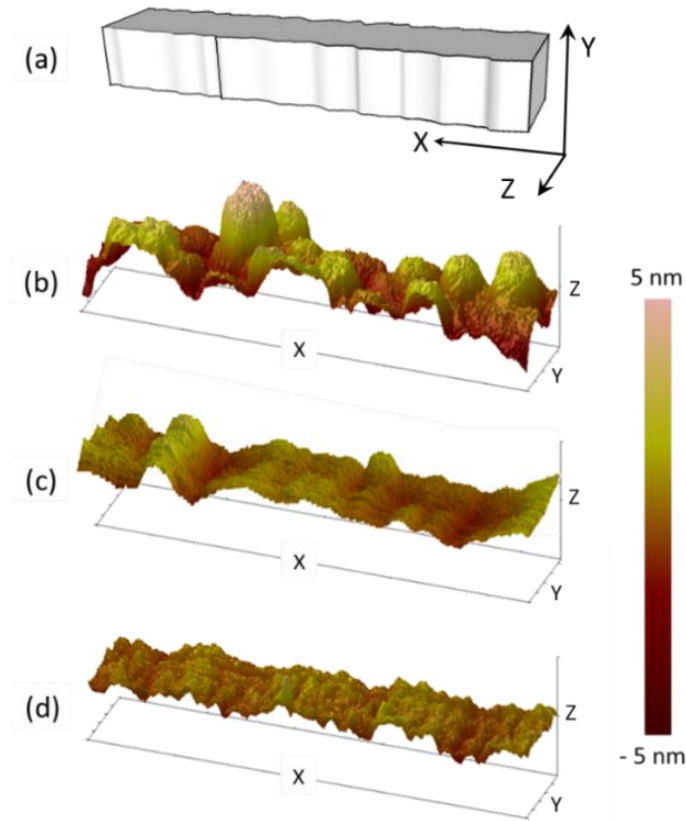

**Figure 4.** 3D image of AFM scans of the waveguide sidewall showing a region 1 µm long by 200 nm high, visually comparing the sidewall roughness of three processes. The scales on all images are equal. (a) Schematic showing the waveguide cross section and the assigned axes of the following AFM images. (b) Process 1: DUV photo-lithography together with a standard $CHF_3/O_2$ etching recipe. (c) Process 2: DUV photo-lithography together with a low-polymer $CHF_3/O_2/N_2$ etching recipe. (d) Process 3: Electron-beam lithography together with a low-polymer $CHF_3/O_2/N_2$ etching recipe.

## 4. Optical Measurements of Quality Factor

It is useful to be able to predict the device performance from a measurement of the sidewall roughness of a waveguide. Even sub-nm roughness can significantly influence the optical performance of a waveguide device. The quality factor is a reliable way to estimate optical performance and correlate the roughness measured by AFM with the losses that we measure in the fabricated device. To estimate device performance, we fabricated ring resonators with a fixed radius of 115 µm and varying widths using e-beam lithography together with the low-polymer $CHF_3/O_2/N_2$ etching recipe in $Si_3N_4$. This fabrication method was chosen because (as described above) it has the lowest sidewall roughness and enables fabrication of the highest quality state-of-the-art photonic devices.

We use a laser scanning technique to measure the transmission spectra and the linewidth of the resonator (FWHM) to estimate the Q of a resonator[33]. The relationship between FWHM and Q is shown in **Equation 3 and Equation 4**. $Q_L$ is the loaded quality factor and $Q_i$ is the intrinsic quality factor. The on-resonance (normalized) transmission minimum $T_{min}$ is between 0 and 1 and the plus (+) and minus (−) signs correspond to the under- and over-coupled regimes respectively.

$$Q_L = \frac{\omega}{\Delta\omega_{FWHM}} = \frac{\lambda}{\Delta\lambda_{FWHM}} \quad (3)$$

$$Q_i = \frac{2Q_L}{1\pm\sqrt{T_{min}}} \quad (4)$$

We launch light from a tunable laser source, then transmit it through a fiber polarization controller and couple it into our device via an inverse nanotaper[34] using a lensed fiber. We collect the light output from the resonator through another inverse nanotaper and an objective lens, which projects the output onto a high-speed InGaAs photodetector. We show the schematic of the experimental setup in **Figure 5**. The laser frequency is measured using a wavemeter with a precision of 0.1 pm, and the laser detuning is calibrated by monitoring the fringes of a reference fiber-based Mach-Zehnder interferometer with a known free spectral range (FSR)[35,36].

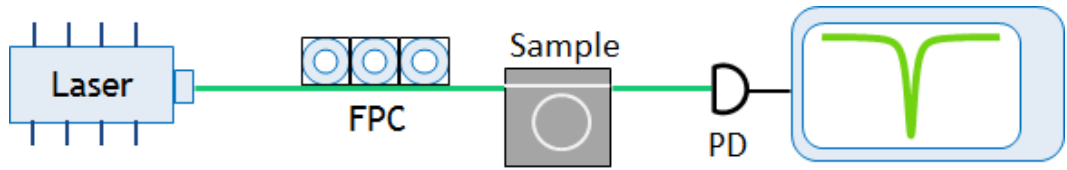

**Figure 5.** Schematic of the experimental setup for measuring transmission spectra and the linewidth of the resonator to characterize device's quality factor (Q) and the propagation loss. FPC indicates the fiber polarization controller and PD indicates the photodetector.

Using the described optical methods, we measured the quality factor of ring resonators with a fixed radius of 115 µm and varying widths which were fabricated together on the same wafer. **Table 2** shows the measured intrinsic Qs for ring resonators with widths of 1200 nm, 1600 nm, 2000 nm, 2400 nm, 5000 nm and 10000 nm. The standard deviations are measurement statistics calculated from hundreds of sweeps on the same ring and include random errors from the measurement setup and analysis noise in the data. These values are also validated using multiple rings. In **Figure 6**, we show four examples of measured normalized TE transmission spectra of ring resonators with widths of 1200 nm, 2000 nm, 2400 nm, and 10000 nm. We see that as the waveguide widths increase the Q also increases, which corresponds to decreasing propagation losses.

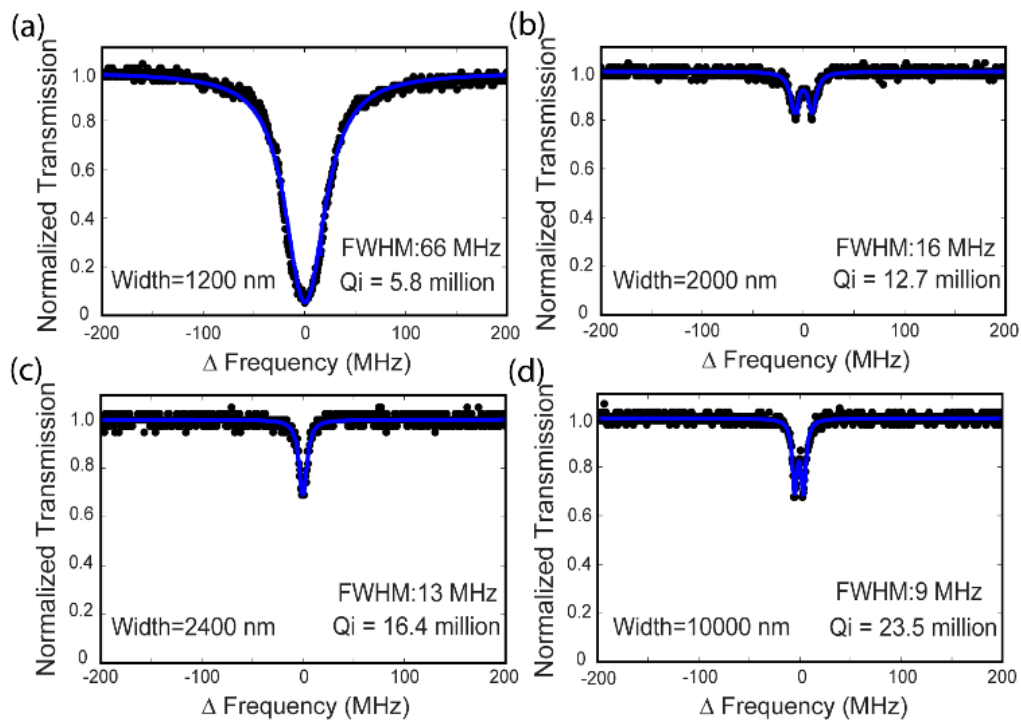

**Figure 6.** Examples of measured normalized TE transmission spectra of ring resonators with different widths fabricated on the same wafer using the same process. Ring radius is fixed at 115 µm. (a) Measured full-width-half-maximum (FWHM) linewidth of 66 MHz for a waveguide width of 1200 nm, which corresponds to intrinsic Q of 5.8 million. (b) Measured FWHM linewidth of 16 MHz for a waveguide width of 2000 nm, which corresponds to intrinsic Q of 12.7 million. (c) Measured FWHM linewidth of 13 MHz for a waveguide width of 2400

nm, which corresponds to intrinsic Q of 16.4 million. (d) Measured FWHM linewidth of 9 MHz for a waveguide width of 10000 nm, which corresponds to intrinsic Q of 23.5 million.

**Table 2.** Quality factors and losses for ring resonators with a fixed radius of 115 µm and varying waveguide width, measured at a wavelength of 1560 nm.

| Width (nm) | 1200 | 1600 | 2000 | 2400 | 5000 | 10000 |
|---|---|---|---|---|---|---|
| Qi (million) | 5.8 ± 0.5 | 8.1 ± 1.3 | 12.5 ± 1.9 | 16.1 ± 3.0 | 22.6 ± 3.2 | 22.9 ± 4.3 |
| Propagation Loss (dB/m) | 5.02 ± 0.38 | 3.60 ± 0.60 | 2.34 ± 0.37 | 1.81 ± 0.35 | 1.29 ± 0.19 | 1.27 ± 0.25 |

**5. Loss Calculations**

The relationship between the quality factor and the sidewall roughness measured by AFM is correlated via the loss calculations. The propagation loss of a ring resonator ($\alpha_{ring}$) in terms of quality factor Q can be written as (**Equation 5**)[33,37]:

$$\alpha_{ring} = \frac{2\pi n_g}{Q_i \cdot \lambda_0} = \frac{\lambda_0}{Q_i \cdot R \cdot FSR} \quad (5)$$

In this equation, $\lambda_o$ is the wavelength of light, $FSR$ is the Free Spectral Range, $Q_i$ is the quality factor, $n_g$ is the group index and $R$ is the radius of the resonator.

The total loss in a ring resonator ($\alpha_{ring}$) can be expressed in terms of two loss mechanisms as follows (**Equation 6**):

$$\alpha_{ring} = \alpha_{surface\,scatter} + \alpha_{bulk} \quad (6)$$

The $\alpha_{bulk}$ represents the losses from the bulk material which includes not only absorption loss due to specific chemical bonds, but also losses from impurities, internal defects and nanovoids. The $\alpha_{surface\_scatter}$ term represents the surface scattering loss which can be related to waveguide geometry and surface roughness parameters by the widely used Payne-Lacey model[38] for losses in a waveguide given by:

$$\alpha_{surface\,scatter} = 4.34 \frac{\sigma^2}{\sqrt{2} k_0 d^4 n_1} g \cdot f \quad (7)$$

Where $\alpha_{surface\_scatter}$ is the waveguide scattering loss in dB per unit length, $\sigma$ is the roughness, $k_o$, $d$, and $n_1$, are the free-space wave vector, the waveguide half width, and the core index, respectively. The function $g$ is determined purely by the waveguide geometry, while $f$ is determined by the correlation length and index step of the waveguide. Details of the expressions $g$ and $f$ can be found in the Payne-Lacey model[38](See Supporting Information). The Payne-Lacey model directly correlates the roughness measurements to the device's optical loss. However, this model was developed to successfully predict loss in straight waveguides, and losses realized in waveguides with bends or resonators are not considered in the standard Payne-Lacey model[38].

## 6. Payne-Lacey Bending Model for Propagation Loss Prediction

We modify the Payne-Lacey model by adding a correction factor which allows us to accurately predict the loss of devices with resonators or bent waveguides given a value of sidewall roughness. The loss caused by scattering depends on the magnitude of the overlap between the optical mode and the waveguide sidewall, where the fabrication imperfections are present.

In **Figure 7**, we can see that in the bent waveguide the light is significantly shifted towards the outer edge of the waveguide in comparison to a straight waveguide, a feature which dramatically increases the sidewall interaction and subsequently increases the scattering loss. This is an effect which the standard Payne-Lacey model represented by **Equation 7** does not incorporate.

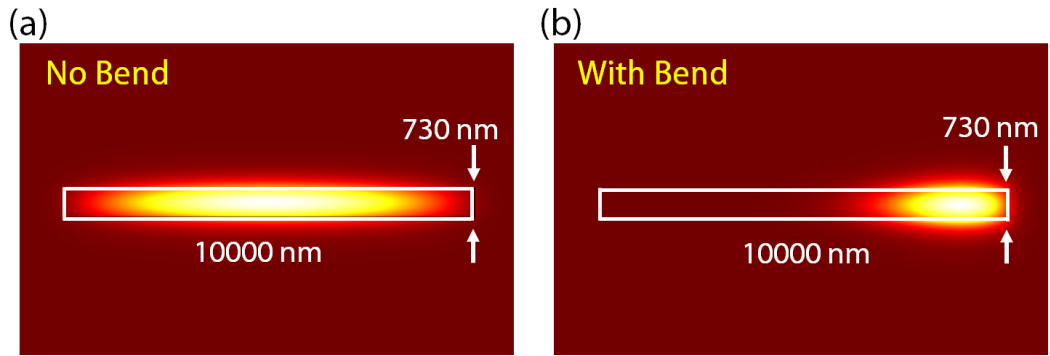

**Figure 7.** Mode simulation for waveguide that is 10000 nm wide and 730 nm high using $Si_3N_4$ as the core material and $SiO_2$ as the cladding material with and without taking bending into account. (a) TE mode profile of a straight waveguide with no bend. (b) TE mode profile of waveguide with a bending radius of 115 μm. Even though a bending radius of 115 μm is already relatively large for an integrated photonic device, we can see the mode overlap with the sidewalls is dramatically different with and without bend.

In order to accurately account for this increased interaction, we add a multiplicative correction factor $\eta$ to the Payne-Lacey expression. The modified Payne-Lacey Bending (PLB) model is expressed as follows in **Equation 8**:

$$\alpha_{surface\_scatter} = 4.34 \frac{\sigma^2}{\sqrt{2}k_0 d^4 n_1} g \cdot f \cdot \eta \quad (8)$$

The correction factor $\eta$ is the ratio of the mode overlap of the bending mode field over the mode overlap for the straight waveguide geometry, a multiplicative factor which gives the magnitude of the increased interaction. This mode overlap is found by integrating the mode field distribution over the waveguide sidewall surface and it is calculated using a finite element method simulation (performed with COMSOL) for each of the bending radii. For straight waveguides $\eta$ in the Payne-Lacey Bending model is taken to be 1. Here we calculate correction factor $\eta$ for the fundamental mode. Higher order modes interact more strongly with the sidewalls and a different $\eta$ must be calculated for each mode.

In order to illustrate the necessity of the correction factor, we plot the Payne-Lacey relationship of sidewall scattering losses both with and without the added correction factor for a ring resonator waveguide with a 10000 nm width and a 115 µm bend radii in **Figure 8**. The measured intrinsic Q was 22.9 ± 4.3 million measured at a wavelength of 1560 nm, and with Equation 5 we find total propagation losses of 1.27 ± 0.25 dB/m. We separate out the loss contributions using Equation 6. We first apply the standard Payne-Lacey model to calculate the scattering losses from the top and bottom surfaces of the waveguide. Using the surface roughness and method reported in previous work, we find the loss to be 0.47 dB/m and 0.24 dB/m for the top and bottom surface respectively[1]. We then calculate the upper and lower bounds of the sidewall scattering loss to be 0.577 dB/m and 0.096 dB/m respectively. (The detailed steps of these calculations and the values used in our case can be found in Supporting Information). The red solid lines in **Figure 8** and the shaded area in between show the Payne-Lacey prediction for roughness and correlation length using the original Payne-Lacey model without the correction factor (Equation 7), while the blue solid lines and the shaded area in between are showing the Payne-Lacey Bending model with the correction factor (Equation 8) for the same resonator with the same measured losses. Meanwhile, the data point on the graph marks the measured roughness and correlation length with error bars obtained from the AFM analysis for the same fabrication process (e-beam lithography with low polymer etching) that we used to fabricate these rings. This data point indicates that, given the measured roughness and correlation length, Payne-Lacey Bending model (with the correction factor) can accurately predict the loss; the traditional model is less accurate and its predicted roughness is much higher than that measured by AFM.

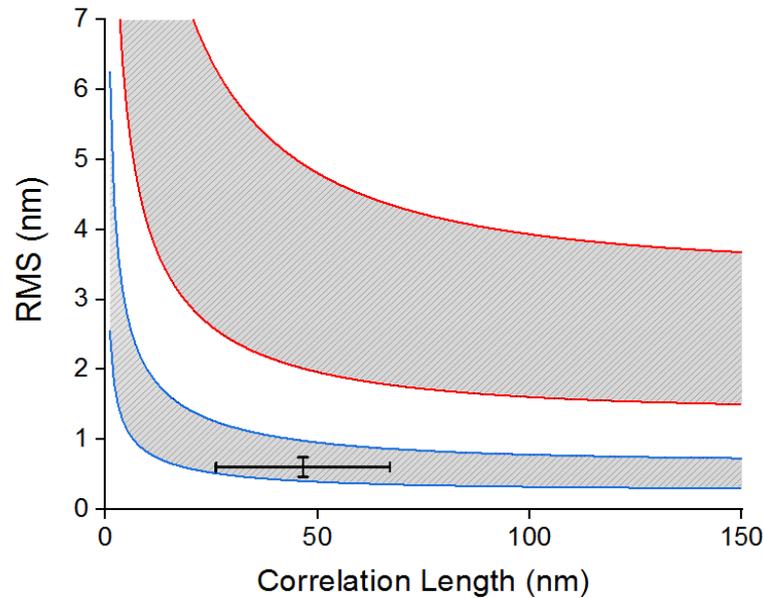

**Figure 8**. Contour map of the sidewall scattering loss as a function of roughness and correlation length for waveguides of 10000 nm width using the model without the correction factor. The red solid lines represent the upper and lower bound of the measured sidewall scattering loss using the model without the correction factor, and the blue solid lines represent the upper and lower bound of the measured sidewall scattering loss using the model with the correction factor. The crossing black lines on the graph mark the range of the measured roughness and correlation lengths obtained from the AFM analysis for the same e-beam and low-polymer etching process that we used to fabricate these rings.

The magnitude of the correction factor changes as a function of both the bending radii and the waveguide width, which is illustrated in **Figure 9(a-b)**. We see the correction factors increase with decreasing bend radius, which physically reflects the mode being increasingly pushed towards the outer sidewall as the bend radius decreases, thereby increasing the magnitude of the mode-sidewall interaction. Meanwhile, **Figure 9(a-b)** also illustrates that the correction factor increases with increasing waveguide width, which can be understood when we consider the ratio that makes up the correction factor. For narrow waveguides, the mode sidewall overlap with the sidewalls is high for the straight waveguide case, and the magnitude of this sidewall overlap does not change significantly even when the waveguide is bent, leaving the ratio of the two close to 1. However, as the waveguide becomes wider, the mode does not see significant overlap with the sidewalls until bending occurs, and this is when the ratio of the mode overlap between the bent and the straight waveguide becomes more dramatic. Therefore, we see in general that the correction factor (proportional to losses) is the largest in the case of small bending radius and large waveguide width, which accurately reflects the high propagation losses seen in integrated microdisks with small bending radii (tens of microns or even hundreds of microns).

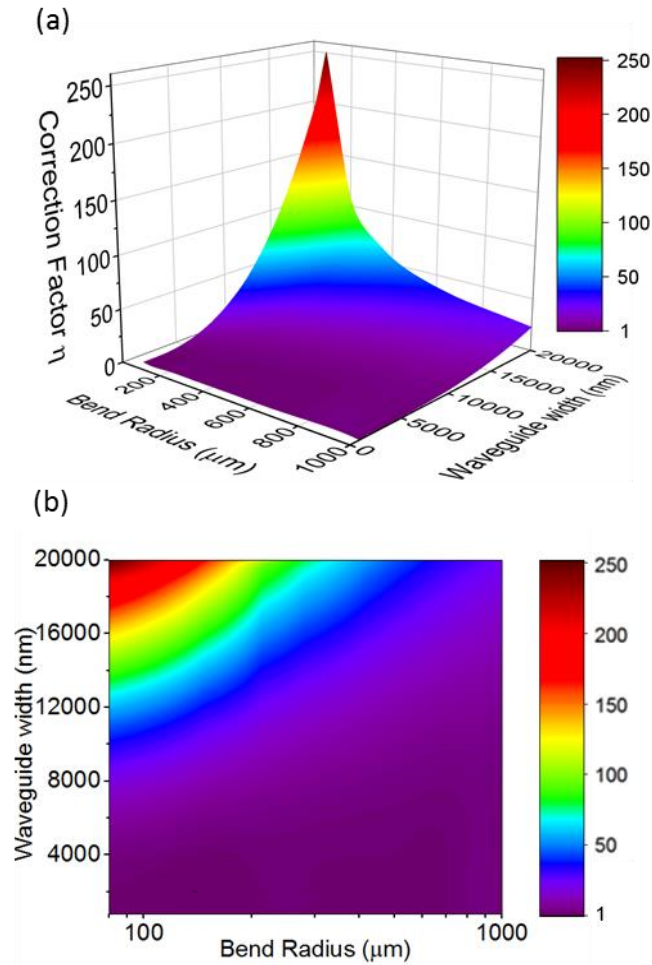

**Figure 9.** The correction factors for various waveguide widths and bend radius. (a) 3D plot of correction factors for various waveguide widths and bend radius. (b) 2D contour map of correction factor vs waveguide widths and bend radius. The correction factors are increasing with decreasing bend radius and increasing waveguide width. When the bending radius is small and the waveguide width is large, the correction factor is the largest, because the mode interaction with the sidewall is more asymmetrical.

To demonstrate the effectiveness of our new Payne-Lacey Bending model, we use the modified Payne-Lacey Bending model to calculate the contribution from the sidewall scattering loss for each of the waveguide widths and plot the results in **Figure 10**. The data points represent experimental loss obtained from the quality factor measurements, and the solid red line shows the theoretical loss calculated by our new Payne-Lacey Bending model for TE mode using the RMS roughness of 0.53 nm and correlation length of 47.4 nm as was measured by AFM.

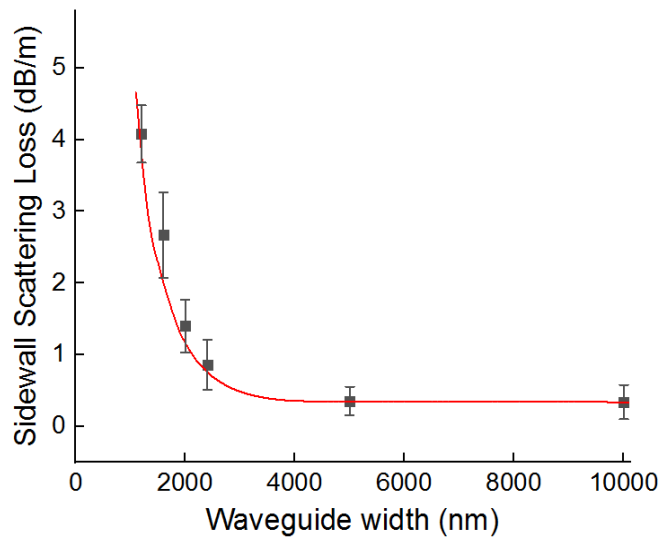

**Figure 10.** Relationship between sidewall scattering loss for TE mode and waveguide width. Data points represent experimental loss, while the solid line shows theoretical loss calculated by our new Payne-Lacey Bending model using a wavelength of 1560 nm. The theoretical losses calculated by the Payne-Lacey Bending model are in good agreement with our experimental results.

We can see that there is an excellent fit between our experimental data and the Payne-Lacey Bending model. We recall that the modified theory only takes into account the percent increase in the overlap between the optical mode and the sidewall for each waveguide width. The fact that this theory is accurate for a range of waveguide widths shows that this model correctly identifies the scattering losses attributed to the sidewall roughness (as opposed to bulk losses and top and bottom scattering losses). The excellent fit also reflects the accuracy of the sidewall roughness measurement made by direct AFM imaging.

## 7. Conclusion and Discussion

In conclusion, we have developed a method which allows us to directly measure sub-nm sidewall roughness of photonic devices using standard AFM. This is a robust method that can be used to characterize the sidewall roughness of waveguides made of any material with sub-nm resolution. Using the new method, we report the sidewall roughness measurement of state-of-the-art $Si_3N_4$ photonic devices for the first time, and compare the influence of e-beam lithography versus DUV photo-lithography as well as the roughness contributions of two different nitride etching recipes. We have further developed a Payne-Lacey Bending model to predict scattering losses. By measuring the losses of waveguide devices with various widths, we experimentally show that it can successfully predict the sidewall scattering losses for devices with resonators or bent waveguides. The combination of this AFM measurement method with the new Payne-Lacey Bending model is very useful for researchers and engineers from both academia and industry to characterize and optimize fabrication processes.

The results of these measurements show there is room for pushing the limits of the new state-of-the-art devices. We have shown that the e-beam lithography provides smoother sidewalls when compared with DUV photo-lithography, with a reduction in the roughness of more than a factor of 2 with all other processes being equal. Currently, our DUV photo-lithography mask is written by a direct laser mask writer. However, in the future, in order to achieve even smoother sidewalls, one could write the DUV photo-lithography mask using e-beam lithography and then project the patterns onto wafer using a DUV or the EUV stepper, where the roughness in the mask pattern will undergo a reduction due to the optics of the stepper. As Equation 7 and 8 show, losses decrease as the square of any reduction in the roughness, $\sigma$. Therefore, the propagation loss in state-of-the-art photonic devices could be further reduced and the DUV stepper would also help realize massively parallel fabrication with a lower cost as well.

Compared with SEM or confocal microscopy imaging where the sidewall roughness is calculated using images, our method of directly using AFM to measure the sidewall roughness yields a greater amount of information. Not only does it allow us to achieve sub-nm resolution, but it also provides us with surface information on the entire sidewall area instead of just one line cut. For example, from line cuts in Figure 3, it can be seen that the roughness at the bottom of the waveguide is smoother than that at the top of the waveguide by 0.21 nm; this feature cannot be distinguished by other imaging methods. The smoother roughness at the bottom of the waveguide could be due to the shorter exposure time in the plasma during etching, and this could be useful for designing a structure that confines modes towards the bottom of the waveguide to reduce the sidewall scattering loss. Additionally, from the 3D data shown in Figure 4, we can also see that the roughness along the X-axis is different from that along the Y-axis, with the X-axis having a larger scale anisotropic roughness which can be attributed to the lithographic patterning and the Y-axis having an isotropic roughness introduced by the etching. Specifically, our measurements of the average vertical (Y-axis) RMS roughness is only 0.31 nm. This analysis allows us to evaluate the individual contributions of each step of the fabrication and to concentrate efforts of sidewall roughness reduction on the steps that have the greatest impact on the roughness values.

**Supporting Information**

**Published in *Advanced Optical Materials*. DOI: 10.1002/adom.202102073 (2022).**

Supporting Information is available from the Wiley Online Library for Adv. Optical Mater., DOI: 10.1002/adom.202102073

**Acknowledgements**

The authors would like to acknowledge Meredith Metzler for helpful discussion regarding fabrication methods and thank Michele McHugh and Gay Miller for assistance in manuscript preparation. Research reported in this work was performed in part at the Cornell NanoScale Science & Technology Facility (CNF), a member of the National Nanotechnology Coordinated Infrastructure (NNCI) supported by National Science Foundation (Grant NNCI-2025233). The authors acknowledge support from the Defense Advanced Research Projects Agency (HR0011-19-2-0014), the Air Force Office of Scientific Research (FA9550-15-1-0303, FA8650-19-C-1002), and the National Science Foundation (OMA-1936345).